\documentclass[amsmath, amsthm, amssymb, floatfix, reprint, onecolumn, notitlepage, nofootinbib, 12pt]{revtex4-1}
\usepackage{setspace}
\usepackage{amsmath}
\usepackage{breqn}
\usepackage{graphicx}
\usepackage[nearskip,margin = 0pt,caption=false]{subfig}

\usepackage{verbatim}
\usepackage{amsfonts}
\usepackage{amssymb}
\usepackage{epstopdf}
\usepackage{lipsum}
\usepackage{siunitx}
\usepackage{xcolor}
\usepackage{array}
\usepackage{braket}
\usepackage[normalem]{ulem}
\usepackage[resetlabels,labeled]{multibib}
\DeclareGraphicsExtensions{.pdf,.eps,.png,.jpg,.mps}
\usepackage{etoolbox}
\usepackage{subfiles} 
\renewcommand{\baselinestretch}{2.0}

\DeclareUnicodeCharacter{0327}{ }

\begin{document}
\title{Integrated Passive Nonlinear Optical Isolators
}
\baselineskip=12pt
\author{Alexander D. White$^{1, \dagger, *}$, Geun Ho Ahn$^{1, \dagger, *}$, Kasper Van Gasse$^{1}$,  Ki Youl Yang$^{1}$, Lin Chang$^{2}$, John E. Bowers$^{2}$, Jelena Vu\v{c}kovi\'{c}$^{1}$\\
\vspace{+0.05 in}
$^1$E. L. Ginzton Laboratory, Stanford University, Stanford, CA 94305, USA.\\
$^2$ Department of Electrical and Computer Engineering, University of California, Santa Barbara, 93106, California, USA\\
{\small $^{\dagger}$These authors contributed equally to this work.}\\
{\small $*$ adwhite@stanford.edu, gahn@stanford.edu}}
\baselineskip=24pt

\begin{abstract}
    \renewcommand{\baselinestretch}{2.0}
    \noindent \textbf{
    Fiber and bulk-optical isolators are widely used to stabilize laser cavities by preventing unwanted feedback. However, their integrated counterparts have been slow to be adopted.
    While several strategies for on-chip optical isolation have been realized, these rely on either integration of magneto-optic materials or high frequency modulation with acousto-optic or electro-optic modulators. 
    Here, we demonstrate an integrated approach for passively isolating a continuous wave laser using the intrinsically non-reciprocal Kerr nonlinearity in ring resonators. 
    Using silicon nitride as a model platform, we achieve single ring isolation of 17-23\,dB with 1.8-5.5\,dB insertion loss, and a cascaded ring isolation of 35\,dB with 5\,dB insertion loss.
    Employing these devices, we demonstrate hybrid integration and isolation with a semi-conductor laser chip.
    }
\end{abstract}

\maketitle
\renewcommand{\baselinestretch}{2.0}

\newpage

\section{Introduction}

The effort to integrate high-performance optical systems on-chip has made tremendous progress in recent years. Advancements in ultra-low-loss photonic platforms \cite{xiang2021laser}, nonlinear photonics \cite{kippenberg2018dissipative}, and heterogenous material integration \cite{de2021iii, xiang2021laser} have enabled fully integrated turnkey frequency comb sources \cite{shen2020integrated, xiang2021laser}, on-chip lasers with Hertz linewidth \cite{jin2021hertz}, Tbps communications on-chip \cite{yang2021inverse, shu2022microcomb}, on-chip optical amplifiers \cite{liu2022_edwa}, and much more. While these systems will continue to improve, a lack of integrated optical isolation limits their performance.


Optical isolators allow for the transmission of light in one direction while preventing transmission in the other. This non-reciprocal behavior is critical in optical systems in order to stabilize lasers and reduce noise by preventing unwanted back-reflection \cite{jalas2013and}. In traditional fiber and bulk optical systems, the non-reciprocal transmission is achieved via use of Faraday effect induced non-reciprocal polarization rotation under an external magnetic field \cite{jalas2013and, srinivasan2018magneto, du2018monolithic}. This approach can be replicated on-chip by integrating magneto-optic materials into waveguides \cite{srinivasan2018magneto}. However, the scalability of the approach still remains a significant challenge due to the custom material fabrication and lack of complementary metal–oxide–semiconductor (CMOS) compatibility. Furthermore, the magneto-optic materials require a very strong magnet for their operation due to their weak effects in the visible to near-IR wavelength range \cite{bi2011chip, yan2020waveguide}, and are therefore difficult to operate in an integrated platform. 

More recently, there has been remarkable progress in integrating magnet-free isolators using an active drive to break reciprocity. This drive has taken the form of a synthetic magnet \cite{tzuang2014non, fang2017generalized}, stimulated Brillouin scattering \cite{kim2015non, kittlaus2018non}, and spatio-temporal modulation 
\cite{tian2021_sin_magnetfree, kittlaus2021electrically_yale, sohn2021electrically}. However, the requirement of an external drive increases the system complexity, often requires additional fabrication, and consumes power. 
Additionally, high power radio-frequency drives contribute large amounts of electromagnetic background that can interfere with the sensitive electronics and photodetection in photonic integrated circuits.
This poses inevitable challenges to the scalability and adoption of such devices.
Therefore, to maximize the scalability and integration into current photonic integrated circuits, an ideal isolator would be fully passive and magnet-free.

Optical nonlinearity presents a promising path towards this direction \cite{sounas2018broadband, yang2020inverse, hua2016demonstration, del2018microresonator, cao2020reconfigurable} owing to its inherent presence of the nonlinearity in most of the widely utilized photonic platforms such as silicon nitride \cite{xuan2016high, kippenberg2018dissipative}, silicon \cite{yang2020inverse}, gallium phosphide \cite{wilson2020integrated}, tantala \cite{jung2021tantala}, silicon carbide \cite{lu2014optical, guidry2020optical} and lithium niobate \cite{lu2019periodically, wang2019monolithic}. 
Unfortunately, due to dynamic reciprocity, many proposals for non-reciptrocal transmission using optical nonlinearities cannot function as isolators  \cite{shi2015limitations}. However, by carefully choosing the mode of operation, isolation using optical nonlinearity is possible and has been demonstrated with discrete components \cite{del2018microresonator}.




Here, we demonstrate integrated continuous-wave isolators using the Kerr effect present in thin-film silicon nitride ring resonators. The Kerr effect breaks the degeneracy between the clockwise and counterclockwise modes of the ring and allows for nonreciprocal transmission. These devices are fully passive and require no input besides the laser that is being isolated. As such, the only power overhead is the small insertion loss through coupling of the ring resonator. Additionally, many integrated optical systems that would benefit from isolators already have high quality silicon nitride or commensurate components and could easily integrate this type of isolator with CMOS compatible fabrication \cite{xiang2021laser}.

By varying the coupling of the ring resonators, we can trade-off insertion loss and isolation. As two examples, we demonstrate devices with a peak isolation of 23\,dB with\,4.6 dB insertion loss and isolation of 17\,dB with a 1.3\,dB insertion loss with 90\,mW of optical power. As we are using an integrated photonics platform, we can reproducibly fabricate and cascade multiple isolators on the same chip, allowing us to demonstrate two cascaded isolators with an overall isolation ratio of 35\,dB. Finally, we butt-couple a semiconductor laser diode chip to the silicon nitride isolators and demonstrate optical isolation in a system on a chip.

\section{Theory of Operation}

The Kerr effect is the change in refractive index of a material due to its third order nonlinearity in susceptibility, $\chi^{(3)}$. In the presence of two electric fields, the nonlinear polarization corresponding to this term is given by $P^{(3)}(t) = \epsilon_0 \chi^{(3)} (E_1 e^{-i\omega_1 t} + E_2 e^{-i\omega_2 t})^3$. Expanding this polynomial and keeping only the terms with the same frequencies, we find that $P^{(3)}(\omega_1) = 3 \epsilon_0 \chi^{(3)}(|E_1|^2 + 2|E_2|^2 )E_1 e^{-i\omega_1 t}$ and $P^{(3)}(\omega_2) = 3 \epsilon_0 \chi^{(3)}(2|E_1|^2 + |E_2|^2 )E_2 e^{-i\omega_1 t}$. Thus, there is an effective increase in the refractive index proportional to the optical intensity. Critically, the index change differs by a factor of 2 depending on the source of the optical power. The field that is degenerate with the mode in consideration contributes a refractive index increase of $3 \epsilon_0 \chi^{(3)}|E|^2$, self-phase modulation (SPM). The field that is non-degenerate contributes a refractive index increase of $6 \epsilon_0 \chi^{(3)}|E|^2$, cross-phase modulation (XPM). This difference provides an intrinsic non-reciprocity. If a strong pump beam is sent through a waveguide, and a weak probe is sent through in the other direction, the probe will accrue an additional phase shift due to the Kerr effect twice that of the pump.

\begin{figure}[h!]
\centering\includegraphics[width=0.95\linewidth]{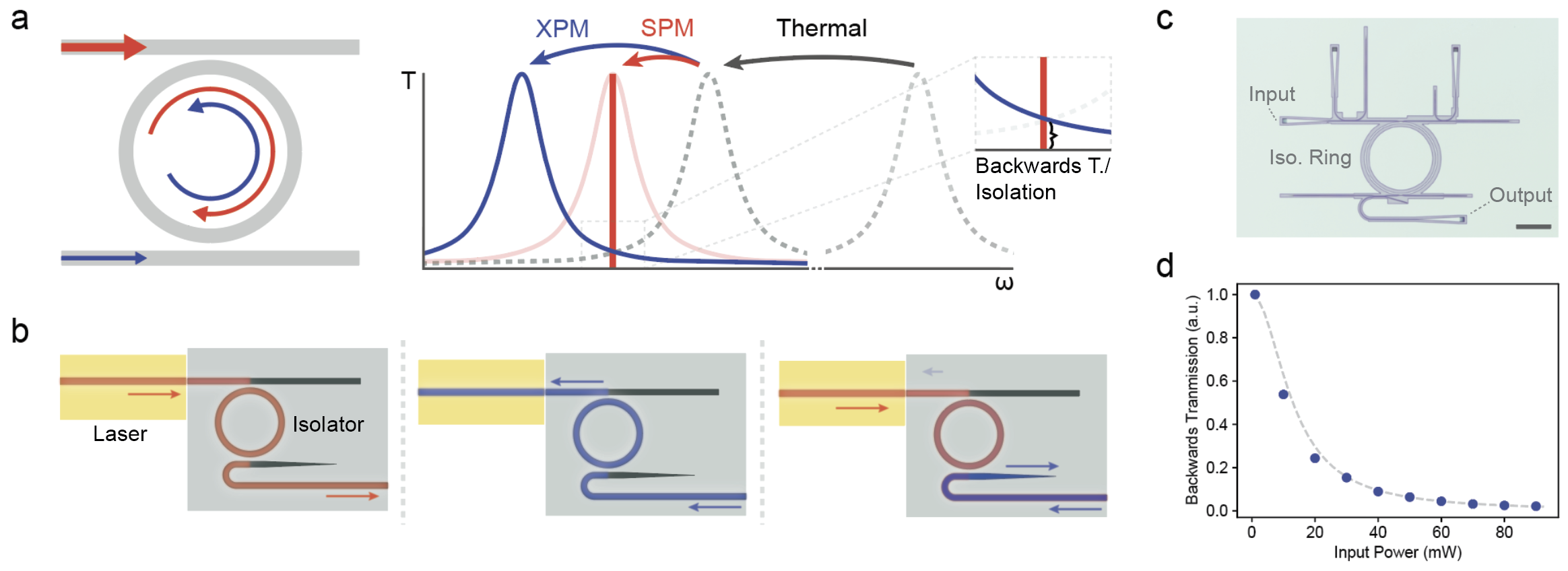}
\caption{{\bf Theory of Operation.} {\bf a}. Schematic showing the operation principle of the integrated nonlinear optical isolators. 
{\bf b}. Illustration of isolator coupled directly to a laser that drives it: in the presence of the laser only (red), unwanted backward transmission only (blue), and laser with backward transmission. When the laser is on, the backward transmission is no longer resonant and the laser is isolated. {\bf c}. Image of silicon nitride device. Scale bar is 100 $\mu m$. {\bf d}. Theoretical (dashed line) and experimental (blue data points) backwards transmission with varied input pump power, illustrating Lorentzian transmission shape.}
\label{fig:Fig1}
\end{figure}

We can apply the same principle to construct an isolator. Consider the setup shown in Figure 1a. A strong pump (red) is sent through a ring resonator with degenerate clockwise and counterclockwise resonances. This pump heats the ring, leading to a reciprocal thermo-optic increase in refractive index and corresponding decrease in resonance frequency. Additionally, the high power in the ring leads to a self-phase modulation of the clockwise mode and a cross-phase modulation of the counterclockwise mode. This shifts the resonance of the counter-clockwise mode twice as far as the clockwise pump mode. The now-split resonances allow for a near unity transmission in the pump direction but significantly reduce the transmission at the same frequency in the reverse direction (blue). This reduction is given by the Lorentzian lineshape of the cavity. Following reference \cite{del2018microresonator}, we can calculate the expected isolation by combining this transmission reduction with the SPM resonance shift:
\begin{equation}
    I = \frac{1}{1 + (2Q\frac{\Delta\omega}{\omega_0})^2,
}
\end{equation}
where the shift $\Delta\omega$ is given by
\begin{equation}
    \Delta\omega = \omega_0 \frac{n_2}{n} \frac{Q \lambda}{2\pi V_{\text{mode}}}  \eta P_{\text{in}},
\end{equation}
where Q is the loaded quality factor of the ring, $n_2$ is the nonlinear refractive index, $n$ is the linear refractive index, $V_{\text{mode}}$ is the mode volume of the ring, and $\eta$ is the coupling efficiency of the pump to the ring. We can characterize the power required for isolation by considering the input power required to isolate by 3\,dB. We will refer to this power level as the isolation threshold, $P_{\text{thresh}}$, given by
\begin{equation}
    P_{\text{thresh}} = \frac{n}{n_2}\frac{\pi V_{\text{mode}}}{Q^2\lambda \eta}
\end{equation}

This isolation, unaffected by dynamic reciprocity (Supplemental Section 1), is achieved solely by the intrinsic non-reciprocity of the ring, and thus there is no additional power required for operation. Additionally, it is important to note that this isolation ratio holds true not only for backwards propagating signals with powers that are small compared to the pump, but even for backwards signals commensurate to and stronger than the pump. As the backwards wave is not resonant with the cavity, the required input power to negate the mode spliting is in fact many times higher than the power of the pump \cite{del2017symmetry}. 

While the bandwidth of the isolation is limited by the resonance splitting, it is possible to add an additional linear filter that indefinitely extends the isolation bandwidth (Supplemental Section 2). Without this additional filter, the 3\,dB bandwidth of the isolation can be given by:
\begin{equation}
    \omega_{3\text{dB}} = 2\Delta\omega - \sqrt{2}\sqrt{\Delta\omega^2 - \frac{\omega_0^2}{4Q^2}},
\end{equation}
which is on the order of the linewidth of the cavity and grows as the isolation increases.

As this type of isolator requires continuous pump power (either with a CW pump or pump that is pulsed at the ring free spectral range), but no additional driving or modulation, it is ideal for directly isolating the output of a laser, Fig 1b. The laser itself acts as the sole driver of isolation, and the device incurs no power consumption, losing power only to the small insertion loss from traversing the ring. There is no need for strong magnetic fields, active optical modulation, or high power RF drives, and device operation is not limited to a single photonic platform or wavelength range.

\section{Device integration and measurement}

As the isolation devices depends on $Q^2$, the mode volume, the nonlinear refractive index, and the input power, it is critical to implement devices with a material that can support high quality microresonators, has an appreciable $\chi^{(3)}$, and can handle very high optical intensities without incurring loss. 
Here, we demonstrate the integrated isolators using silicon nitride as a model system, as it has become one of the most prominent platforms for integrated nonlinear photonics \cite{xiang2021laser}.



Here we use thin film silicon nitride ($<$ 400 nm), as it has the potential for CMOS integration compatibility given the lower film stress present \cite{moss2013new, kim2017dispersion}. In addition, the thin silicon nitride process allows for geometric dispersion properties that easily lead to a strong normal dispersion \cite{kim2017dispersion}, allowing us to suppress spurious optical parametric oscillation (Supplemental Section 3). 
To maximize $Q^2/V_{\text{mode}}$ while keeping the isolator compact, we use a ring diameter of 200~$\mu$m, as shown in Fig 1c.


\begin{figure}[h!]
\centering\includegraphics[width=0.95\linewidth]{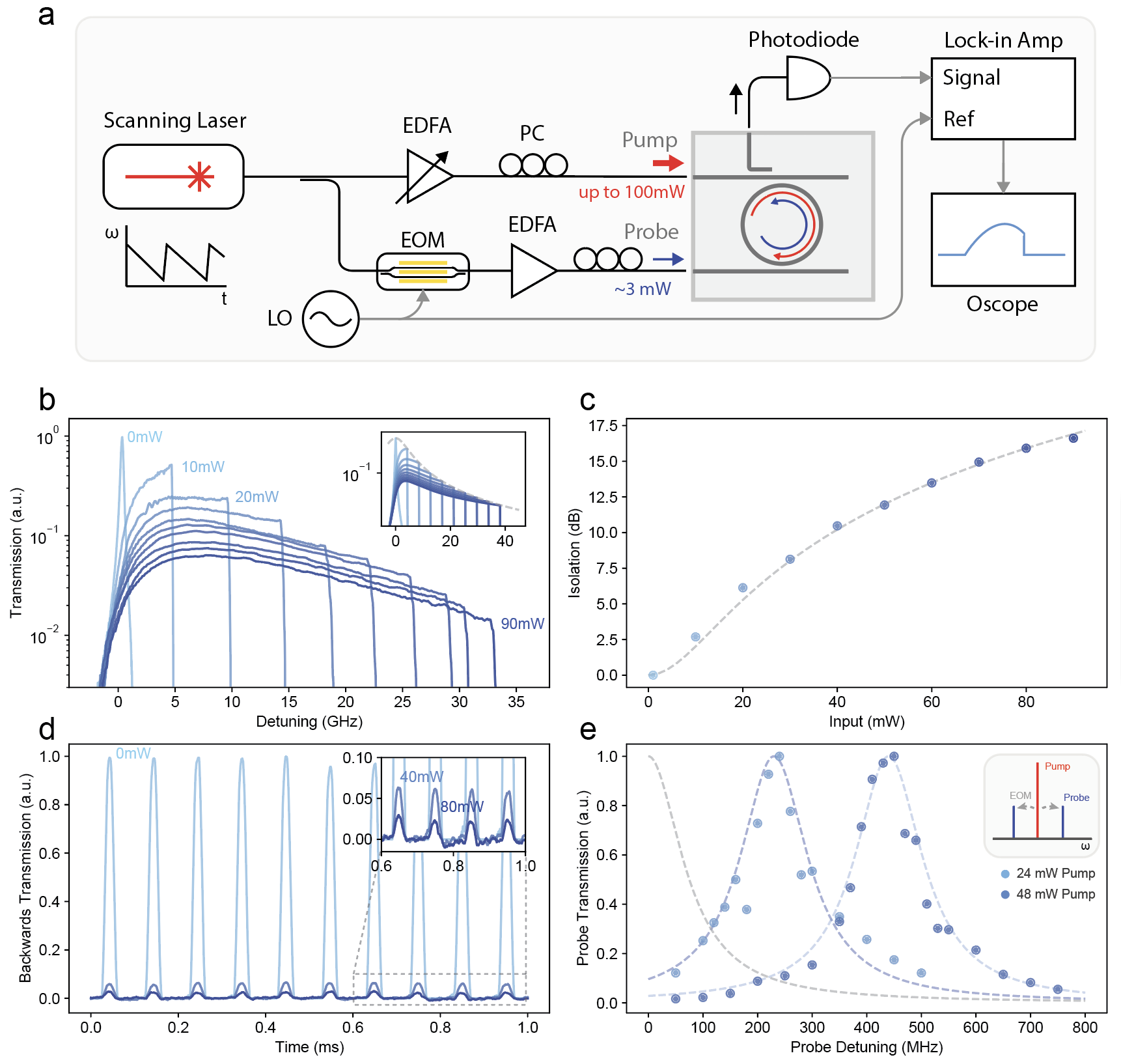}
\caption{{\bf Isolation Measurement} {\bf a}. Schematic of the measurement set up for characterizing the nonlinear optical isolators. PC: polarization controller. LO: 90\.KHz electronic oscillator. {\bf b}. Pump power-dependent measurement of backwards transmission. Inset shows the theoretical pump power dependence. {\bf c}. Corresponding device isolation. {\bf d}. Pulsed backward transmission measurement with increasing pump power (0\,mW, 40\,mW, 80\,mW). Inset shows magnification. {\bf e}. Frequency dependence of backwards transmission. Here, the probe is split into two sidebands with an EOM, and this sideband separation is swept with a frequency synthesizer. As expected, the backwards frequency response is shifted proportionally to the pump power.
}
\label{fig:Fig2}
\end{figure}

To measure the isolation of these devices, we use the pump-probe setup illustrated in Fig 2a. As the pump and probe are sourced from the same laser, they have the same optical frequency. For the first set of measurements, shown in Figs. 2b-c, the pump and probe wavelengths are scanned across the ring resonance, while for Fig. 2d the pump is kept fixed. We send a high power pump through the ring and simultaneously modulate and send a low power probe through the ring in the opposite direction. We then scan the pump and probe across the resonance and read the reverse transmission using a lock-in amplifier. During the scan, the pump thermally pulls the ring until the ring unlocks at the peak of its resonance. By monitoring the probe transmission at the resonance peak, we can get a direct measurement of the isolation (Supplemental Section 4). Additionally, by varying the pump power we can measure the power dependent isolation, Fig. 2b-c. As the pump power is increased, the peak isolation is red shifted and scales as a Lorentzian. We find excellent agreement between our measurements (Fig 2b) and the expected transmission from a simple model of a thermally pulled ring with a Lorentzian power dependent isolation (Fig. 2b inset). 

We also validate the operation of the isolator with a static pump frequency. The ring remains locked to the laser and we can directly measure the backwards transmission of the device by sending optical pulses at the same frequency as the pump, Fig 2d. Finally, we can measure the frequency response of the isolation by modulating the probe
using an electro-optic modulator (EOM). This generates sidebands that we can sweep across the resonance. As only the red-shifted sideband will be resonant with the red-shifted backwards resonance, we can sweep the sideband frequency to map out the frequency response, Fig 2e. We find, as expected from the XPM modulation that the backwards transmission has a Lorentzian profile detuned from the pump by the SPM resonance shift $\Delta\omega$.

\begin{figure}[h!]
\centering\includegraphics[width=0.95\linewidth]{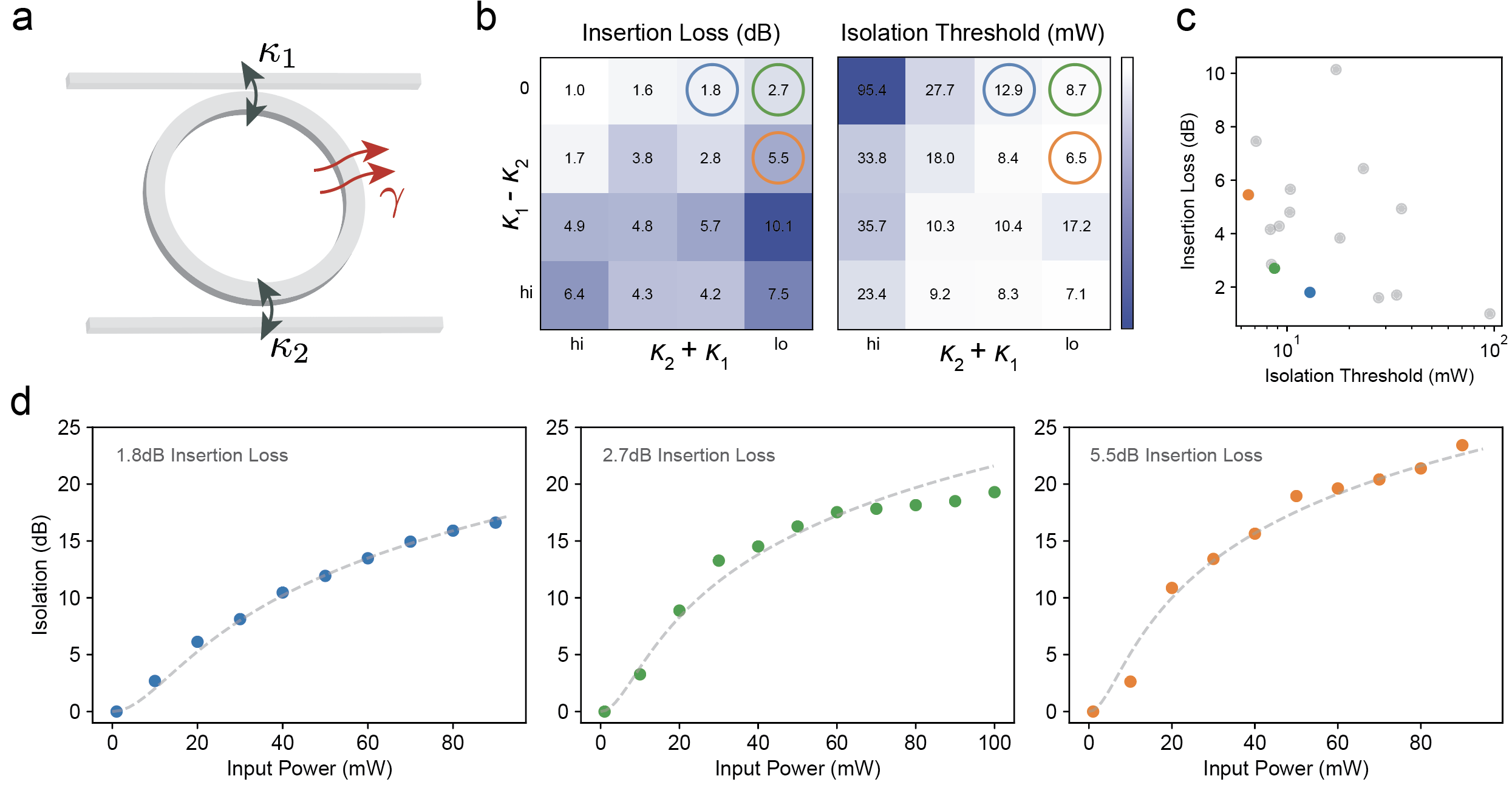}
\caption{{\bf Performance Optimization} {\bf a}. Schematic of isolator ring illustrating key parameters $\kappa _1$, $\kappa _1$, and $\gamma$, 
the input coupling rate, output coupling rate, and intrinsic loss rate, respectively.
{\bf b}. Heatmaps showing measured insertion loss and isolation threshold for varied coupling rates $\kappa _1$ and $\kappa _2$. Color bar limits set by min and max of each plot. Well-performing parameters are highlighted with blue, green and orange circles, and {\bf c}. Correlations of the isolation and insertion losses from b.  {\bf d}. Pump power-dependent isolation for the three highlighted rings.
}
\label{fig:Fig3}
\end{figure}

To maximize the performance of these isolators, it is important to consider both insertion loss and isolation. In this system, these are determined by the coupling rate to the two waveguides, $\kappa_1$ and $\kappa_2$, and the scattering rate of the ring into the environment, $\gamma$. Ideally, all power is transmitted into the ring, and all of the power in the ring is transmitted to the output port. This is possible by increasing the ring coupling rates, but this has the effect of reducing the Q of the resonance and thus lowering the isolation. To maximize the isolation, the power must be efficiently transferred to the ring, but the coupling rates should be minimized as to preserve the Q. This, of course, increases insertion loss. 
More precisely, the ring sees a power of $\frac{4\kappa_1(\kappa_2 + \gamma)}{(\kappa_1 + \kappa_2 + \gamma)^2}$, the Q is impacted by a factor of $\frac{1}{\kappa _1 + \kappa _2 + \gamma}$, and the insertion loss is given by $\frac{4 \kappa_1 \kappa_2}{(\kappa_1 + \kappa_2 + \gamma)^2}$.

To interrogate this trade-off experimentally, we fabricate an array of 16 air-clad silicon nitride isolators with varying coupling strengths and coupling asymmetries, Fig 3b-c. We find these devices have an intrinsic quality factor of approximately 5 million (Supplemental Section 5). As expected, devices with weaker and more asymmetric coupling show higher isolation, but also higher insertion loss. We highlight the performance of two of the devices; a device with 1.8\,dB insertion loss and an isolation threshold of 12.9\,mW, and a device with 5.5\,dB insertion loss and an isolation threshold of 6.5\,mW (Fig 3d). These devices show a peak isolation at 90\,mW of 16.6\,dB and 23.4\,dB respectively.

As these isolators are integrated and can have low insertion loss, it is possible to fabricate and cascade multiple devices on the same chip, enabling an exponential enhancement in isolation, Fig 4a. To test this, we fabricate two rings, the second slightly red-detuned from the first. This allows for the thermal shift to bring both rings onto resonance and lock them there. 
The isolation is maximized and overall insertion loss minimized at a given pump power when the second ring is red detuned by a factor of the single ring insertion loss times the thermal pulling of the first ring (Supplemental Section 6). To characterize the isolation of cascaded rings, we first measure the power dependent isolation of a single ring, Fig 4c, using the same pump probe measurement as described in Fig 2a. We then repeat this measurement for two cascaded rings, one slightly red-detuned from the second. These results are shown in figures 4d and 4e.
The multiplicative effect of the cascaded rings enables us to achieve an isolation of 35\,dB with an insertion loss of approximately 5\,dB. 

\begin{figure}[t!]
\centering\includegraphics[width=0.95\linewidth]{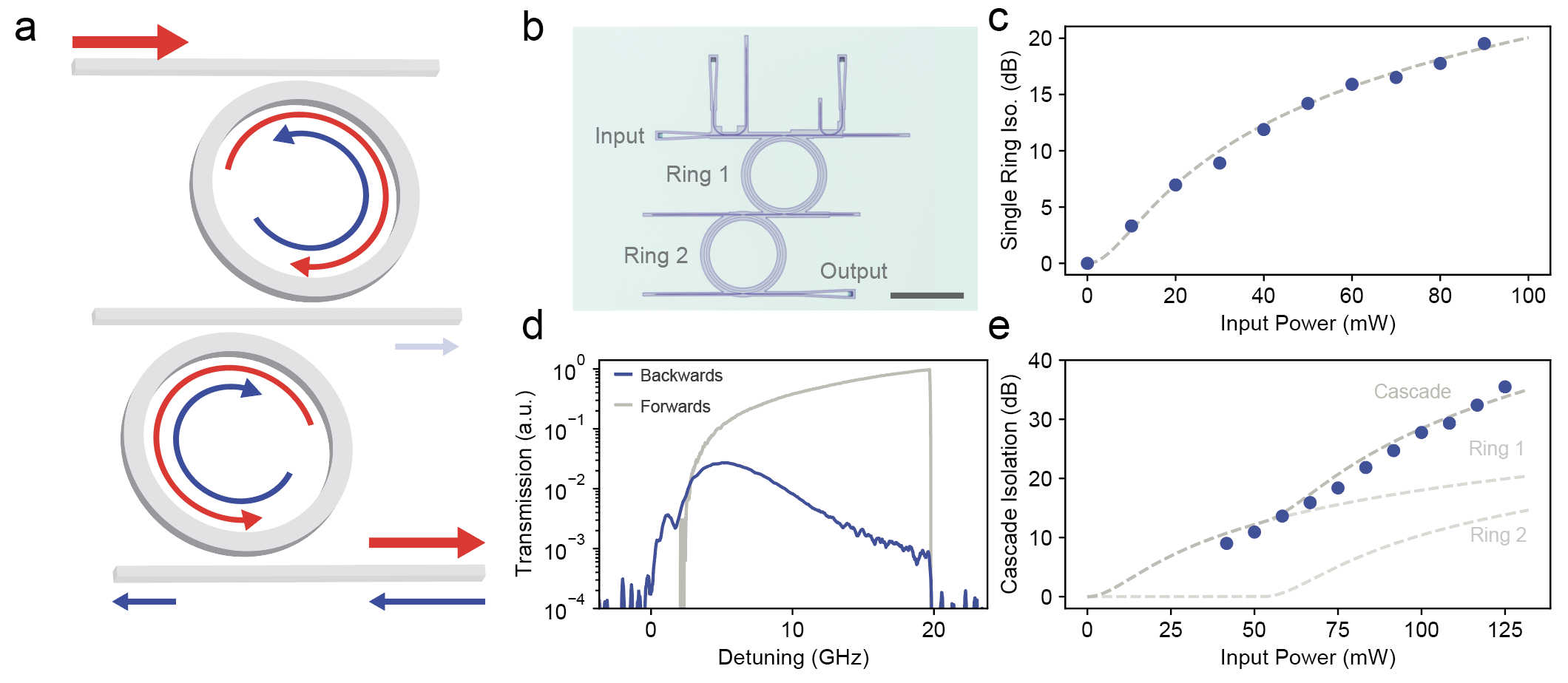}
\caption{{\bf Isolator Cascade} {\bf a}. Schematic of cascaded isolator rings. {\bf b}. Optical micrograph of fabricated cascaded isolator rings. Scale bar is 200\/$\mu m$ {\bf c}. Power-dependent single ring isolation. {\bf d}. Transmission in the forwards and backwards direction from the cascaded isolator rings with a 110mW pump.  {\bf e}. Power dependant isolation of cascaded rings. Theoretical fit calculated by multiplying the isolation ratio from a single ring to a second ring red-shifted from the first. Measurements start from 40~mW as this much pump power is needed to overlap the two ring resonances.
}
\label{fig:Fig5}
\end{figure}

\begin{figure}[h!]
\centering\includegraphics[width=0.95\linewidth]{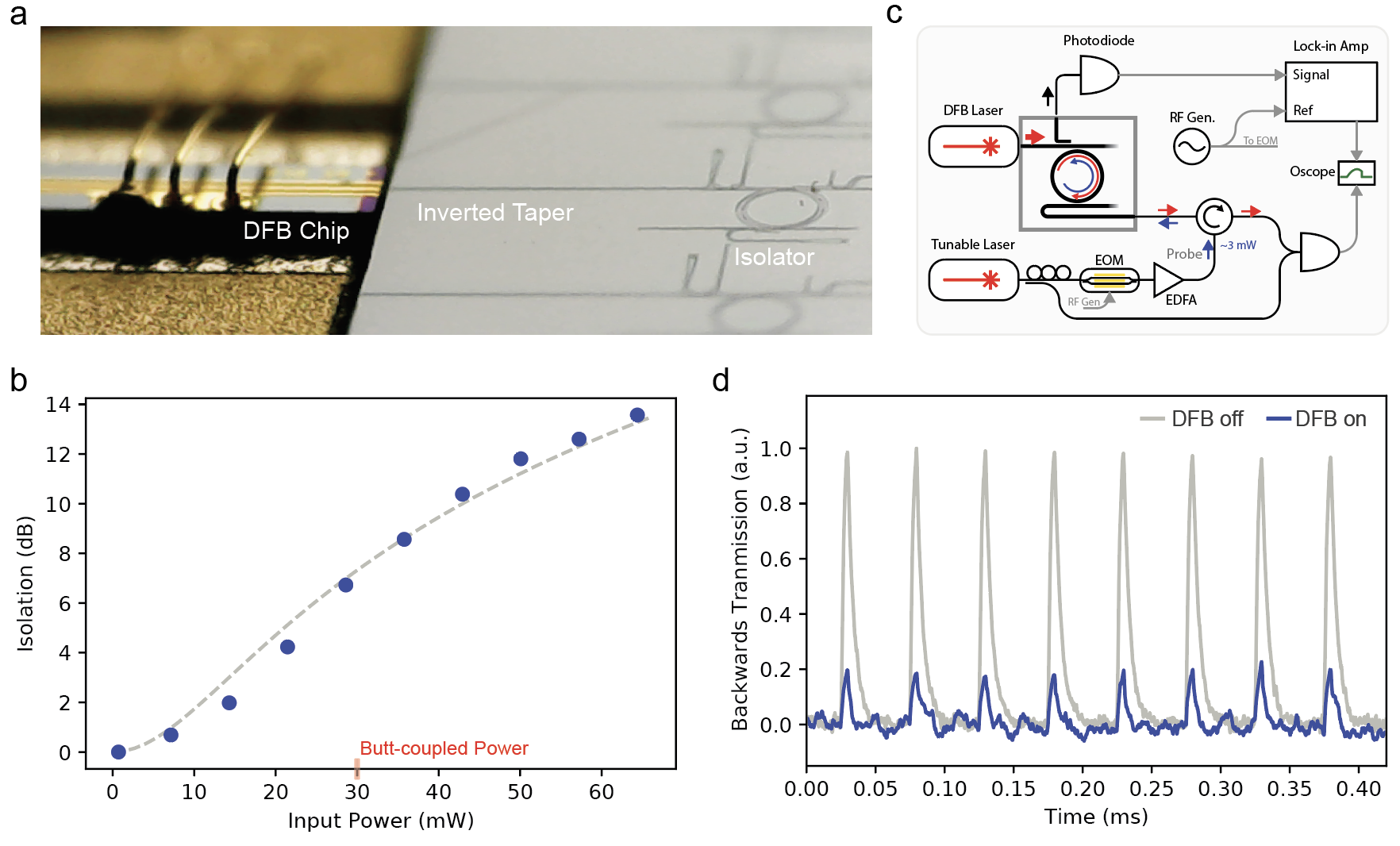}
\caption{{\bf DFB Hybrid Integration} {\bf a}. Optical image of hybrid integration of DFB laser with the isolator. {\bf b}. Power dependent isolation measured with the amplified DFB laser. {\bf c}. Schematic of experimental measurement set up for direct measurement of hybrid integrated DFB-isolator operation. {\bf d}. Transmission of backwards pulses with the directly coupled DFB laser on and off. 
}
\label{fig:Fig4}
\end{figure}  

Finally, we demonstrate isolation using a DFB laser chip, Fig 5a. To maximize the on-chip pump power, we couple the DFB laser to the chip using an oxide-clad inverted taper designed to match the output mode of the laser \cite{shen2020integrated}. We first characterize the isolation by coupling the DFB laser to a lensed fiber and performing a pump-probe measurement similar to Fig 2a. To tune the DFB laser across the ring resonance we modulate its temperature using a peltier and thermistor for feedback. We observe isolation up to 13.6\,dB with 65\,mW input power (Fig 5b), slightly lower than before due to a small reduction in Q factor. We then directly butt-couple the DFB laser and isolator, and thermally lock the ring to the laser. To verify its isolation, we send pulses backward through the device using a secondary laser and measure their transmission, Fig 5c-d. To ensure the secondary laser is at the same frequency as the DFB, we mix the laser outputs on a photodiode and minimize their beat-tone.

\section{Conclusion}

We demonstrate on-chip optical isolators utilizing the Kerr effect that are fully passive. By tuning the coupling parameters we trade-off between insertion loss and isolation, demonstrating devices with insertion loss of only 1.8\,dB with 17\,dB isolation, and single ring isolation of up to 23\,dB. Due to the integrated nature of these isolators, they can be easily cascaded to improve performance. By cascading two rings, we achieve 35\,dB isolation with 5\,dB insertion loss. Finally, we demonstrate the application of such a device to isolate the output of an edge-coupled DFB laser chip. 

As these devices are fully passive and magnet-free, they require no external drive and can operate without generating any electromagnetic interference or magnetic field background.
In spite of this, their performance is still competitive with state-of-the-art active and magnetic integrated isolators (Table S1) \cite{tian2021_sin_magnetfree, sohn2021electrically, kittlaus2021electrically_yale, yan2020waveguide, du2018monolithic, herrmann2021mirror, lira2012electrically, dostart2021optical, fang2017generalized}. Furthermore, better controlled fabrication from commercial foundries will allow for higher quality factors \cite{liu2021high} and enable cascading of more than two rings, pushing the power threshold for 20\,dB isolation down to below 2\,mW and the achievable isolation to over 70\,dB (Supplemental Section 8). As many hybrid- and heterogeneously-integrated optical systems already contain high quality photonics in Kerr materials, this type of isolator can be immediately incorporated into state of the art integrated photonics.

\bibliography{References}

\clearpage 

\section*{Methods}

\noindent\textbf{Device Fabrication} \\
The thin film silicon nitride (310 nm) is deposited on silicon dioxide/silicon carrier wafer using low-pressure chemical vapor deposition (LPCVD). The isolator device patterns are defined using JEOL JBX-6300FS electron-beam lithography, using with ZEP520A as the electron resist. Post-development, the patterns are transferred to silicon nitride by inductively coupled plasma (ICP) etching with CHF${_3}$/CF${_4}$ chemistry. After the etch, resist is removed using piranha solution, and the silicon nitride chips are subsequently annealed in N${_2}$ environment at 1100 \textdegree C. 

\noindent\textbf{Isolator Measurements}
A scanning laser (Toptica) is split into two paths with a directional coupler. One path serves as the pump and one as the probe. The pump path is passed through a polarization controller and amplified by an EDFA (IPG) before being sent to the chip. The probe path is modulated with an EOM (Optilab), passed through a polarization controller, and amplified by an EDFA (Thorlabs) before being sent to the chip. The backwards transmission is measured using a photodiode (Thorlabs), lock-in amplifier (Stanford Instruments), and oscilloscope (Rigol). Inverse-designed grating couplers optimized for transmsission at 1550nm are used to couple to and from the chip. To minimize leak through from the probe input fiber to the detection fiber, the grating inputs are oriented perpendicular to each-other. For fixed and scanning measurements of power-dependent isolation, the EOM is modulated with a 90kHz signal from an arbirtary waveform generator (Rigol) the same signal is used for lock-in detection. For frequency-dependent measurements, the EOM is driven with an amplified (Minicircuits) 90kHz lock-in signal mixed (Minicircuits) with a high frequency modulation from a frequency synthesizer (Rohde and Schwarz).

\noindent\textbf{DFB Operation}
The DFB laser is driven with a precision source (Kiethley) with 380\,mW electrical power. To thermally stabilize and tune the frequency of the laser, the laser mount is cooled by a peltier using a 10k$\Omega$ thermistor to provide feedback control with a temperature controller (Thorlabs).

\noindent\textbf{Acknowledgments}\\
\noindent 
The authors thank David Carlson, Travis Briles, Jizhao Zang, Jennifer Black, Su-Peng Yu, Felix M. Mayor,  Jason F. Herrmann, Amir H. Safavi-Naeini, Scott Papp, and Rahul Trivedi for collaboration and discussion, and Lue Wu and Kerry Vahala for assistance with the DFB laser. A.W. acknowledges the Herb and Jane Dwight Stanford Graduate Fellowship (SGF) and the NTT Research Fellowship. G.H.A. acknowledges support from STMicroelectronics Stanford Graduate Fellowship (SGF) and Kwanjeong Educational Foundation. Authors from Stanford acknowledge funding support from DARPA under the LUMOS program. Part of this work was performed at the Stanford Nano Shared Facilities (SNSF)/Stanford Nanofabrication Facility (SNF), supported by the National Science Foundation under award ECCS-2026822.

\clearpage

\pagebreak

\end{document}